\documentclass[journal, twocolumn, 10pt]{IEEEtran}

\usepackage{graphicx}
\usepackage{cite}
\usepackage{amssymb,amsmath}
\usepackage{algorithm}
\usepackage{algpseudocode}
\usepackage{epsfig}
\usepackage{epstopdf}
\usepackage{color}
\usepackage{flushend}
\usepackage{amsthm}

\begin{document}

\title{Spectrum Sharing Opportunities of Full-Duplex Systems using Improper Gaussian Signaling}
%

\pagenumbering{gobble}

\author{{ Mohamed Gaafar, Osama Amin, Walid Abediseid, and Mohamed-Slim Alouini} \\
\small  Computer, Electrical and Mathematical Sciences and Engineering (CEMSE) Division, \\ King Abdullah University of Science and Technology (KAUST), \\ Thuwal, Makkah Province, Saudi Arabia. \\ E-mail: {\{{mohamed.gaafar, osama.amin, walid.abediseid, slim.alouini\}@kaust.edu.sa} \vspace*{-20pt}}
 \thanks{The work of M.-S. Alouini was supported by the Qatar National Research Fund (a member of Qatar Foundation) under NPRP Grant NPRP 5-250-2-087. The statements made herein are solely the responsibility of the authors.}
}

\maketitle

\begin{abstract}
Sharing the licensed spectrum of full-duplex (FD) primary users (PU) brings strict limitations on the underlay cognitive radio operation. Particularly, the self interference may overwhelm the PU receiver and limit the opportunity of secondary users (SU) to access the spectrum. Improper Gaussian signaling (IGS) has demonstrated its superiority in improving the performance of interference channel systems. Throughout this paper, we assume a FD PU pair that uses proper Gaussian signaling (PGS), and a half-duplex SU pair that uses IGS. The objective is to maximize the SU  instantaneous achievable rate while meeting the PU quality-of-service. To this end, we propose a simplified algorithm that optimizes the SU signal parameters, i.e, the transmit power and the circularity coefficient, which is a measure of the degree of impropriety of the SU signal, to achieve the design objective. Numerical results show the merits of adopting IGS compared with PGS for the SU especially with the existence of week PU direct channels and/or strong SU interference channels.
\end{abstract}

\section{Introduction}

Underlay cognitive radio (CR) is a new dimension in wireless communications that holds the promise of increasing the opportunity of spectrum access especially after the explosive growth of the spectrum demand over the past decade \cite{zhao2007survey}. It allows the secondary users (SU) to use the licensed spectrum while causing tolerable interference to the primary users (PU) through power limitation policies. Therefore, it becomes a big challenge to improve the SU coverage and data rate performance while satisfying the PU quality-of-service (QoS). The overall improvements in such systems are mainly governed by the progress achieved in interference channel system design. To manage the interference in such systems, several techniques have been proposed in literature. In this paper, we tackle this problem using Improper Gaussian
signaling (IGS)\cite{zeng2013transmit}.

Designing IGS includes optimizing the transmitted power along with a special parameter that measures the degree of impropriety. Optimal adjusting of these parameters increases the degrees of freedom in interference channel systems and thus improves its performance limits. Recently, IGS has been adopted in underlay CR systems and provided a great promise in improving the SU performance  \cite{lameiro2015benefits, amin2015outage}. In \cite{lameiro2015benefits}, Lameiro \textit{et al.} studied the instantaneous achievable rate of both PU and SU assuming IGS at the SU and proper Gaussian signaling (PGS) at PU. Then, the SU power and the circularity coefficient are adjusted to maximize its rate while achieving the PU QoS. IGS achieves better performance than PGS when the PU is not highly loaded. In \cite{amin2015outage}, Amin \textit{et al.} investigated the outage probability of PU and SU of the same system. Then, they designed the SU IGS parameters, based on average channel state information (CSI), to minimize its outage probability considering a predefined PU QoS. The aforementioned research in \cite{lameiro2015benefits, amin2015outage} considered half-duplex PU and SU. Designing underlay CR to coexist with full-duplex (FD) PU is a challenging design problem and has not been studied before.

In-band FD communications allow both communication nodes to achieve simultaneous transmission in the same frequency band. In underlay CR systems, FD is used in conjugation with cooperative communications to increase both the spectral efficiency and coverage  \cite{kim2012optimal, zhongTOAPPEARperformance}, in addition to improving the sensing procedure \cite{afifi2015incorporating}. The current literature did not consider the coexistence of underlay CR with FD networks due to the self-interference impact at the PU, which reduces the possibility of spectrum sharing\cite{sabharwal2014band}.

In this paper, we explore the opportunity of improving spectrum sharing opportunities with in-band FD systems throughout employing IGS at the SU. The main contributions of this paper are:
\begin{itemize}
\item Share the licensed spectrum of FD PU, where the PU design criterion is based on minimum fixed target rate.
\item Develop a simple optimal algorithm that design the SU IGS parameters represented in the 
power and circularity coefficient to maximize the SU rate while satisfying the PU QoS based on instantaneous CSI.
\item  Investigate through numerical results the benefits that can be reaped by employing IGS for the SU when compared with PGS.  
\end{itemize}
\section{System Model}
\begin{figure}[!t]
\centering
\includegraphics[width=3in]{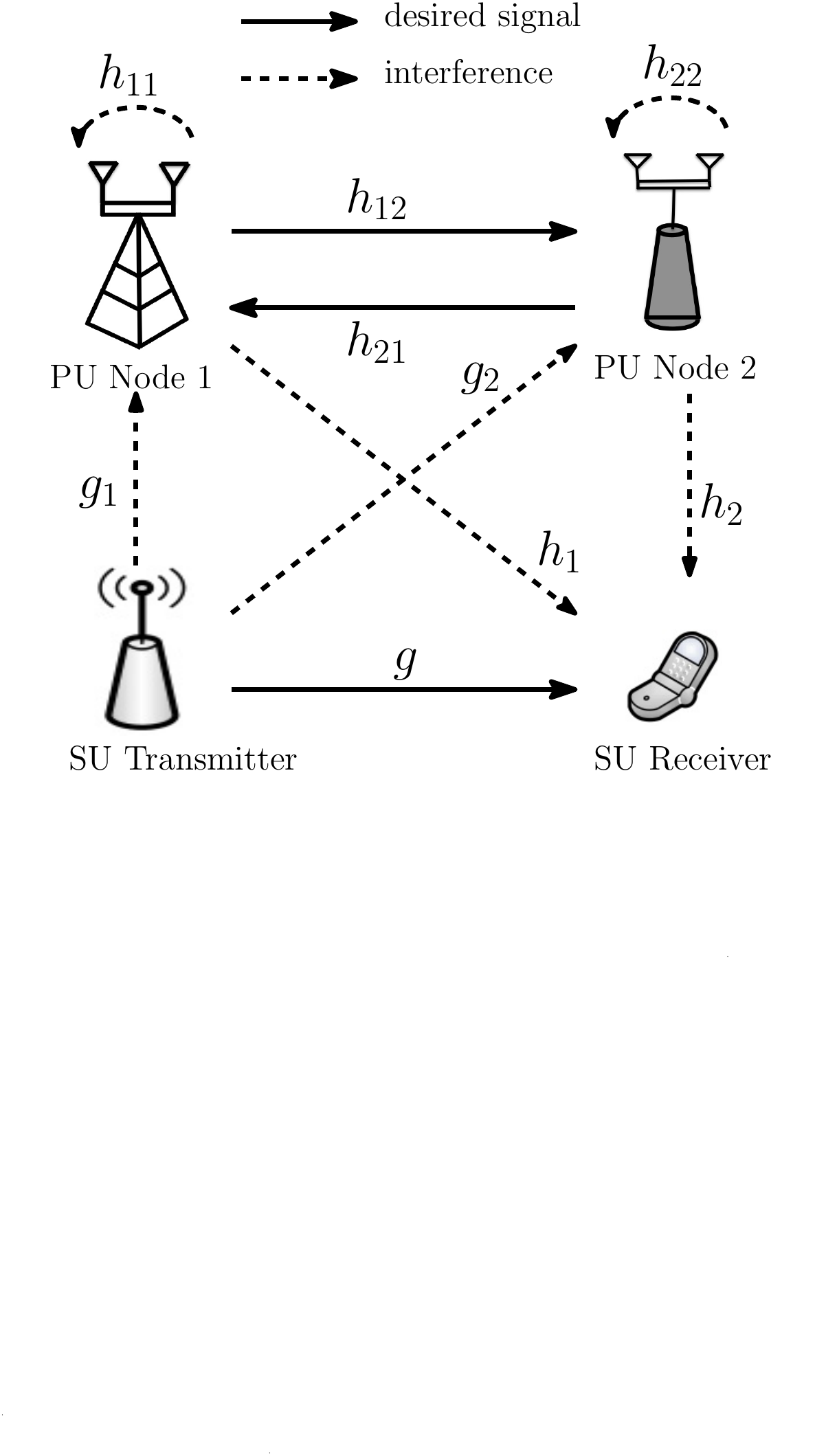}
\caption{System model.}
\label{fig1}
\end{figure}

The spectrum licensed system consists of an in-band FD pair, as shown in Fig. \ref{fig1} that uses zero-mean unity variance PGS $x_i, i \in {\{1,2\}}$. The underlay cognitive system shares the PU spectrum through half-duplex communications and employs IGS $x_\mathrm{s}$ with a unit variance and a circularity coefficient $\mathcal{C}_x$. To clarify the difference between IGS and PGS schemes, we introduce the following definitions: 

\textit{Definition 1}: The variance and pseudo-variance of a zero mean complex random variable $x$ are defined as $\sigma _x^2 = {\mathbb{E}} {{{[\left| x \right|}^2]}} $ and $\bar{\sigma} _x^2 = {\mathbb{E}} {{[{ x }^2]}} $, respectively, where $\mathbb{E}[.]$ is the expectation operator and $\left| . \right|$ is the absolute value \cite{Neeser1993proper}.

\textit{Definition 2}: The \textit{proper} signal has a zero  $\bar{\sigma} _x^2 $, while the \textit{improper} signal has a non-zero $\bar{\sigma} _x^2 $.  

\textit{Definition 3}: The circularity coefficient $\mathcal{C}_x$ is a measure of  the degree of impropriety of signal $x$ and is defined as $\mathcal{C}_x = \left|\bar{\sigma} _x^2 \right|/\sigma _x^2$, where $0 \le {\mathcal{C}_x} \le 1$. $\mathcal{C}_x=0$ denotes \textit{proper} signal and $\mathcal{C}_x=1$ denotes \textit{maximally improper} signal.

The received signal at PU node $j$, where $j \in {\{1,2\}}$, $i \neq j$, is expressed as
\begin{equation} \label{pu_sig}
{y_j} = \sqrt {{p_i}} {h_{ij}}{x_i} + \sqrt {{p_j}} {h_{jj}}{x_j} + \sqrt {{p_{\mathrm{s}}}} {g_{j}}{x_{\mathrm{s}}} + {n_j},  
\end{equation}
where $p_i$ is the transmitted signal power of the PU node $i$, ${p_{\mathrm{s}}}$  is the SU transmitted power, $n_j$ is the additive white Gaussian noise (AWGN) at the receiver of the PU node $j$, $h_{ij}$ denotes the direct communication channel between the PU node $i$ and the PU node $j$, $g_{j}$ represents the interference channel between the SU transmitter and the PU node $j$, and $h_{jj}$ represents the residual self interference (RSI) channel of node $j$ after undergoing analog and digital cancellation techniques. We assume that the RSI is modeled as a zero mean complex Gaussian random variable as in \cite{kim2012optimal, day2012full}. The PU nodes transmit and receive at the same time over the same frequency and hence, we could assume channel reciprocity, i.e., $h_{ij}=h_{ji}$, however this might not always be valid if the PU nodes use different spatial antennas' locations or the receivers' front end and transmitters' back end are not perfectly matched \cite{biglieri2007mimo}. The proposed frame work and algorithm apply to both symmetrical and asymmetrical channels.

On the other hand, the received signal at the SU receiver is written as,
\begin{equation} \label{su_sig}
{y_{\rm{s}}} = \sqrt {{p_{\rm{s}}}} g{x_{\rm{s}}} + \sum\limits_{i = 1}^2 {\sqrt {{p_i}} {h_i}{x_i}}  + {n_{\rm{s}}},
\end{equation}
where $n_\mathrm{s}$ is the AWGN at the SU receiver, $h_{i}$ is the interference channel of the PU node $i$ on the SU receiver, $g$ denotes the direct communication channel between the SU transmitter and receiver. 

The channels in the described system are modeled as Rayleigh flat fading channels  and the additive noise at the receivers end is modeled as a white, zero-mean, circularly symmetric, complex Gaussian with variance~$\sigma^2$. Similar to \cite{lameiro2015benefits}, we assume perfect CSI knowledge, which might be impractical, but it provides beneficial performance bounds of spectrum sharing with FD PU. Furthermore, some  channel estimation implementations are suggested in \cite{Zheng2013Full, Ghasemi2007Fundamental, Zhang2008Optimal}.

By employing IGS, the achievable rate for the PU node $i$ is given by \cite{zeng2013transmit, lameiro2015benefits},
\begin{align}\label{pu_rate_not_simplified}
{R_{{{\mathrm{p}}_i}}}\left( {{p_{\mathrm{s}}},{\mathcal C_x}} \right) = {\log _2}\left( {1 + \frac{{{p_i}{\gamma _{{{\mathrm{p}}_i}}}}}{{ \beta_j + {p_{\mathrm{s}}}{{\mathcal I}_{{{\mathrm{s}}_j}}} }}} \right) + \frac{1}{2}{\log _2} {\frac{{1 - {\mathcal C}_{{{{y}}_i}}^2}}{{1 - {\mathcal C}_{{{\mathcal{I}}_i}}^2}}},
\end{align} 
where ${\gamma _{{{\mathrm{p}}_{_i}}}} = {{{{\left| {{h_{ij}}} \right|}^2}}} / {{{\sigma ^2}}}$ is the channel-to-noise ratio (CNR) of the link from the PU node $i$ to the PU node $j$,  ${{\mathcal I}_{{{\mathrm{s}}_i}}} =  {{{{\left| {{g_{{i}}}} \right|}^2}}}/{{{\sigma ^2}}}$ is the interference CNR of the SU to the PU node $i$, $\beta_j={{p_j}{\upsilon _{{{\rm{p}}_j}}} + 1}$, ${\upsilon _{{{\mathrm{p}}_i}}} = {{{{\left| {{h_{ii}}} \right|}^2}}}/{{{\sigma ^2}}}$ in the RSI CNR at the PU node $i$, ${\mathcal C}{_{{{y}_i}}}$ and ${{\mathcal C}_{{{\mathcal{I}}_i}}}$ are the circularity coefficients of the received and interference-plus-noise signals at PU node $i$, respectively, which are given by
\begin{align}\label{cir_coeff}
{\cal C}{_{{y_i}}} = \frac{{{p_{\rm{s}}}{{\cal I}_{{{\rm{s}}_j}}}{{\cal C}_x}}}{{ {p_i}{\gamma _{{{\rm{p}}_i}}}+\beta_j + {p_{\rm{s}}}{{\cal I}_{{{\rm{s}}_j}}}}}, \hspace{0.2cm} {{\cal C}_{{{\cal I}_i}}} = \frac{{{p_{\rm{s}}}{{\cal I}_{{{\rm{s}}_j}}}{{\cal C}_x}}}{{\beta_j + {p_{\rm{s}}}{{\cal I}_{{{\rm{s}}_j}}} }}.
\end{align}
After some manipulations, ${R_{{{\mathrm{p}}_i}}}\left( {{p_{\mathrm{s}}},{\mathcal{C}_x}} \right)$ can be simplified as 
\begin{align}\label{pu_rate}
{R_{{{\mathrm{p}}_i}}}\left( {{p_{\mathrm{s}}},{\mathcal{C}_x}} \right) = \frac{1}{2} {\log _2}\left( {\frac{{{{\left( {{p_i}{\gamma _{{{\mathrm{p}}_i}}} + \beta_j + {p_{\mathrm{s}}}{{\mathcal I}_{{{\mathrm{s}}_j}}} } \right)}^2} - {{\left( {{p_{\mathrm{s}}}{{\mathcal I}_{{{\mathrm{s}}_j}}}{\mathcal{C}_x}} \right)}^2}}}{{{{\left( {\beta_j + {p_{\mathrm{s}}}{{\mathcal I}_{{{\mathrm{s}}_j}}} } \right)}^2} - {{\left( {{p_{\mathrm{s}}}{{\mathcal I}_{{{\mathrm{s}}_j}}}{\mathcal{C}_x}} \right)}^2}}}} \right).
\end{align}
Similarly, the SU achievable rate can be expressed as
\begin{align}\label{su_rate}
& {R_{\mathrm{s}}}\left( {{p_{\mathrm{s}}},{\mathcal{C}_x}} \right) = \frac{1}{2}{\log _2}\left( {\frac{{p_{\mathrm{s}}^2{\gamma _{\mathrm{s}}}^2\left( {1 - {\mathcal{C}_x^2}} \right)}}{{{{\left( {\sum\nolimits_{i = 1}^2 {{p_i}{{\cal I}_{{{\rm{p}}_i}}}}  + 1} \right)}^2}}} } \right. \nonumber \\ & \qquad \qquad \qquad \qquad \qquad 
\left. {+ \frac{{2{p_{\mathrm{s}}}{\gamma _{\mathrm{s}}}}}{{\sum\nolimits_{i = 1}^2 {{p_i}{{\cal I}_{{{\rm{p}}_i}}}}  + 1}} + 1} \right),
\end{align}
where ${\gamma _{\mathrm{s}}} = {{{{\left| {{g}} \right|}^2}}}/{{{\sigma ^2}}}$ is the SU direct CNR between the SU transmitter and receiver and ${{\mathcal I}_{{{\mathrm{p}}_i}}} = {{{{\left| {{h_{i}}} \right|}^2}}}/{{{\sigma ^2}}}$ is the interference of the PU node $i$ to the SU.

Due to the Rayleigh fading assumption,  the direct, interference and RSI CNR ${\gamma _{{{\mathrm{p}}_{_i}}}}, \, { \gamma _{\mathrm{s}}}, \;{ {\mathcal I}_{{{\mathrm{p}}_i}}}, \; { {\mathcal I}_{{{\mathrm{s}}_i}}}, \, { \upsilon _{{{\mathrm{p}}_i}}}$  are then  exponentially distributed random variables with mean values ${\bar \gamma _{{{\mathrm{p}}_{_i}}}}, \, {\bar \gamma _{\mathrm{s}}}, \, {\bar {\mathcal I}_{{{\mathrm{p}}_i}}}, \, {\bar {\mathcal I}_{{{\mathrm{s}}_i}}}, \, {\bar \upsilon _{{{\mathrm{p}}_i}}}$ respectively.

From \eqref{pu_rate} and \eqref{su_rate} that if $\mathcal{C}_x=0$, we can obtain the well known formulations of the achievable rates of proper signaling as follows
\begin{align}\label{rate_cx_0}
&{R_{\rm{s}}}\left( {{p_{\rm{s}}},0} \right) = {\log _2}\left( {1 + \frac{{{p_{\rm{s}}}{\gamma _{\rm{s}}}}}{{\sum\nolimits_{i = 1}^2 {{p_{_i}}{{\cal I}_{{{\rm{p}}_i}}}}  + 1}}} \right),\nonumber \\ &{R_{{{\rm{p}}_i}}}\left( {{p_{\rm{s}}},0} \right) = {\log _2}\left( {1 + \frac{{{p_i}{\gamma _{{{\rm{p}}_i}}}}}{{{\beta _j} + {p_{\rm{s}}}{{\cal I}_{{{\rm{s}}_j}}}}}} \right).
\end{align}
Moreover, if $\mathcal{C}_x$ increases, the SU rate decreases while the PU rate increases which will allow the SU to increase its transmitted power. Thus, proper adjustment of the SU power and circularity coefficient should be carefully considered to maximize the SU rate and satisfy the PU QoS requirements along with the maximum SU power budget.  
\section{CR Transmitted Signal Design}
In this section, we optimize the SU signal parameters, i.e, transmit power $p_{\rm{s}}$ and circularity coefficient $\mathcal{C}_x$, in order to maximize the SU achievable rate while maintaining a predetermined PU achievable rate constraint for each PU link.
\subsection{Primary Users rates Constraint Based Criterion} 
The $i^{\mathrm{th}}$ PU node design criterion is to achieve a minimum fixed target rate $R_{{\rm{0,}}{{\rm{p}}_i}}$, i.e., $R_{{\rm{p}}_i} \geq {R_{{\rm{0,}}{{\rm{p}}_i}}}$. The PU is assumed to transmit with its maximum power budget $p_i$ to mitigate interference sources such as the RSI, and ensure its QoS requirements. According to the PU perspective, its achievable rate is
\begin{align}\label{pu_rate_design}
{R_{{\rm{p}}_i}} = {\log _2}\left( {1 + \frac{{{p_i}{\gamma _{{{\rm{p}}_i}}}}}{{1 + {{\cal I}_{{{\rm{agg,p}}_i}}}}}} \right),
\end{align}   
where ${{\cal I}_{{{\rm{agg,p}}_i}}}$ is the aggregate interference-to-noise ratio at the $i^{\mathrm{th}}$ PU receiver. As a result, the maximum allowable margin interference-to-noise ratio, ${{\cal I}_{{\rm{max,}}{{\rm{p}}_i}}}$, at the receiver of the $i^{\mathrm{th}}$ PU node  can be found from \eqref{pu_rate_design} at $R_{{\rm{p}}_i} = {R_{{\rm{0,}}{{\rm{p}}_i}}}$ as
\begin{align} \label{Imax}
{{\cal I}_{{\rm{max,}}{{\rm{p}}_i}}} = {\left[ {\frac{{{p_i}{\gamma _{{{\rm{p}}_i}}}}}{{{\Gamma _{{{\rm{p}}_i}}}\left( 1 \right)}} - 1} \right]^ + },
\end{align} 
where $[z]^+=\max(0,z)$ and ${\Gamma _{_i}}\left( x \right) = \left( {{2^{x{R_{{\rm{0,}}{{\rm{p}}_i}}}}} - 1} \right)$ represents the required signal-to-interference-plus-noise ratio (SINR) to achieve a rate of $x{R_{{\rm{0,}}{{\rm{p}}_i}}}$. Hence, according to the instantaneous direct CNR of the PU and its parameters, the PU may be in outage if it cannot attain its required minimum target rate ${R_{{\rm{0,}}{{\rm{p}}_i}}}$. The outage happens if the RSI surpasses the maximum allowable interference. 
\subsection{Proper Gaussian Signaling Design}   
 For the PGS design, the SU allocates its transmit power in order to maximize its achievable rate subject to its own power budget ${p_{{\rm{s,max}}}}$ and PU QoS. As such, we formulate the following optimization problem,
\begin{align}\label{opt_prob_proper}
&\mathop {\max }\limits_{{p_{\rm{s}}}} {R_{\rm{s}}}\left( {{p_{\rm{s}}},0} \right) \nonumber \\
&\mathrm{s.\;t.} \hspace{0.25cm}{R_{{{\rm{p}}_i}}}\left( {{p_{\rm{s}}},0} \right) \ge {R_{{\rm{0,}}{{\rm{p}}_i}}}, \nonumber \\ & \hspace{0.9cm}0 < {p_{\rm{s}}} \le {p_{{\rm{s,max}}}}.
\end{align}
From \eqref{rate_cx_0}, the PU rate constraints in \eqref{opt_prob_proper} reduce to ${p_{\rm{s}}}\leq{p^{\left(i\right)}_{\rm{s}}}$, where ${p^{\left(i\right)}_{\rm{s}}}$ is the 
feasible upper bound of the SU power that satisfies the $i^{\mathrm{th}}$ PU rate constraint and is found to be,
\begin{align}\label{proper_psi}
p_{\rm{s}}^{\left( i \right)} = {\left[ {\frac{{\beta_j}}{{{{\cal I}_{{{\rm{s}}_j}}}}}{\Psi _i}\left( {1,1} \right)} \right]^ + },
\end{align}
where ${\Psi _i}\left( {x,y} \right) = \frac{{{\varphi _i}\left( x \right)}}{{{\Gamma _{_i}}\left( y \right)}} - 1$ and ${\varphi _i}\left( x \right) = {2^{x{R_{{{\rm{p}}_i}}}\left( {0,0} \right)}} - 1$ represents the required signal-to-interference-plus-noise ratio (SINR) to achieve a rate of ${x{R_{{{\rm{p}}_i}}}\left( {0,0} \right)}$ in the absence of the SU. Hence, \eqref{opt_prob_proper} can be rewritten as
\begin{align} \label{opt_prob_proper_equivalent}
&\mathop {\max }\limits_{{p_{\rm{s}}}} {R_{\rm{s}}}\left( {{p_{\rm{s}}},0} \right)  \nonumber \\   &\mathrm{s. \;  t. \;} \; {p_{\rm{s}}} \leq \mathop {\min } \left( p_{\rm{s}}^{\left( 1 \right)}, p_{\rm{s}}^{\left( 2 \right)}, p_{\mathrm{s,max}} \right).
\end{align}
Moreover, from \eqref{rate_cx_0}, one can show that ${R_{\rm{s}}}\left( {{p_{\rm{s}}},0} \right)$ is monotonically increasing in ${p_{\rm{s}}}$, hence the optimal SU transmitted can be found from
\begin{align} 
{p_{\rm{s}}} = \mathop {\min } \left( p_{\rm{s}}^{\left( 1 \right)}, p_{\rm{s}}^{\left( 2 \right)}, p_{\mathrm{s,max}} \right).
\end{align}
Furthermore, one can deduce easily from \eqref{proper_psi} that the SU transmits when the maximum allowable margin interference exceeds the RSI, i.e.,  
\begin{align}\label{working_cond}
{{\cal I}_{{\rm{max,}}{{\rm{p}}_i}}} > {p_{_j}}{\upsilon _{{{\rm{p}}_j}}}.
\end{align}
Otherwise, the SU stays idle.
\subsection{Improper Gaussian Signaling Design}
Different from the previous subsection, the IGS based-system has here an additional parameter, i.e, $\mathcal{C}_x$, which needs to be jointly optimized with $p_\mathrm{s}$ in order to maximize the SU achievable rate under the PU rate constraints. To this end, we formulate the following optimization problem
\begin{align}\label{opt_prob}
&\mathop {\max }\limits_{{p_{\rm{s}}},{\cal C}{_x}} {R_{\rm{s}}}\left( {{p_{\rm{s}}},{\cal C}{_x}} \right) \nonumber \\
&\mathrm{s. \;t.} \hspace{0.25cm}{R_{{{\rm{p}}_i}}}\left( {{p_{\rm{s}}},{\cal C}{_x}} \right) \ge {R_{{\rm{0,}}{{\rm{p}}_{\rm{i}}}}}, \nonumber \\ & \hspace{-0.15cm} \qquad \quad 0 < {p_{\rm{s}}} \le {p_{{\rm{s,max}}}}, \nonumber \\& \hspace{0.9cm} 0 \le {\cal C}{_x} \le 1.
\end{align}  
Unfortunately, this is a non-linear and non-convex optimization problem which makes it hard to obtain its optimal solution. However, thanks to some monotonicity properties of the objective function and the constraints, we get the optimal solution of \eqref{opt_prob} as will be explained in the following.

First, after some manipulations, the PU rate constraints in \eqref{opt_prob} reduce to
\begin{align}\label{ps_improper_quadraric}
{\cal I}_{{{\rm{s}}_j}}^2\left( {1 - {\cal C}_x^2} \right){p_{\rm{s}}^2} - 2{p_{\rm{s}}}{\beta _j}{{\cal I}_{{{\rm{s}}_j}}}{\Psi _i}\left( {1,2} \right) - \beta _j^2{\Psi _i}\left( {2,2} \right) \le 0.
\end{align}
After solving the quadratic inequality in \eqref{ps_improper_quadraric}, we obtain the feasible upper bound of the SU power from
\begin{align}\label{ps_improper_max}
p_{\rm{s}}^{\left( i \right)}\left(\mathcal{C}_x\right)\hspace{-0.1cm} = \hspace{-0.1cm}\left[ {{\beta _j}\frac{{\sqrt {\Psi _i^2\left( {1,2} \right) + \left( {1 - {\cal C}_x^2} \right){\Psi _i}\left( {2,2} \right)}  + {\Psi _i}\left( {1,2} \right)}}{{{{\cal I}_{{{\rm{s}}_j}}}\left( {1 - {\cal C}_x^2} \right)}}} \right]^+ \hspace{-0.1cm}.
\end{align}
Similar to the proper Gaussian signaling design case, one can show easily that if \eqref{working_cond} is valid, then $p_{\rm{s}}^{\left( i \right)}\left(\mathcal{C}_x\right)>0$ and the SU may transmit improper signals, otherwise, it remains silent. Hence, \eqref{opt_prob} can be equivalently reformulated as
\begin{align}\label{opt_prob_mod}
&\mathop {\max }\limits_{{p_{\rm{s}}},{\cal C}{_x}} {R_{\rm{s}}}\left( {{p_{\rm{s}}},{\cal C}{_x}} \right)\nonumber \\
&\mathrm{s. \; t.}\hspace{0.25cm}{p_{\rm{s}}} \le \min \left\{ {p_{\rm{s}}^{\left( 1 \right)}\left( {{{\cal C}_x}} \right), p_{\rm{s}}^{\left( 2 \right)}\left( {{{\cal C}_x}} \right),{p_{{\rm{s}},{\rm{max}}}}} \right\}, \nonumber \\ & \qquad \quad \hspace{-0.2cm} 0\le {\cal C}{_x} \le 1.
\end{align}    
From \eqref{su_rate}, it is clear that $ {R_{\rm{s}}}\left( {{p_{\rm{s}}},{\cal C}{_x}} \right)$ is monotonically increasing in $p_{\mathrm{s}}$ for a fixed ${{{\cal C}_x}}$, thus $p_{\mathrm{s}}$ is assigned the upper bound of the SU power constraint in  \eqref{opt_prob_mod} as
\begin{align}\label{ps_min}
{p_{\rm{s}}} = \mathop {\min } \left( p_{\rm{s}}^{\left( 1 \right)}\left(\mathcal{C}_x\right), p_{\rm{s}}^{\left( 2 \right)}\left(\mathcal{C}_x\right), p_{\mathrm{s,max}} \right).
\end{align}
To solve \eqref{opt_prob_mod}, we obtain the distinct intersection points of the three functions of the SU power constraint in $0 < {{\cal C}_x} < 1$. First, we can show that $p_{\rm{s}}^{\left( i \right)}\left(\mathcal{C}_x\right)$ is strictly\footnote{See Appendix A for the proof} increasing in ${\cal C}_x$ over the interested interval. Thus, the power constraint in \eqref{opt_prob_mod} can be described as a piecewise function with a maximum of four intervals (three breaking points) and a minimum of one interval (zero breaking points). We can calculate the intersection point, $r^{\left( i \right)}$, between $p_{\rm{s}}^{\left( i \right)}$ and ${p_{{\rm{s,max}}}}$ from
\begin{align}\label{intersect_r1_r2}
{r^{\left( i \right)}} = \sqrt {1 - \frac{{\beta _j^2{\Psi _i}\left( {2,2} \right) + 2{p_{{\rm{s,max}}}}{\beta _j}{{\cal I}_{{{\rm{s}}_j}}}{\Psi _i}\left( {1,2} \right)}}{{{{\left( {{p_{{\rm{s,max}}}}{{\cal I}_{{{\rm{s}}_j}}}} \right)}^2}}}},
\end{align}     
which exists  if $p_{\mathrm{s}}^{\left( i \right)}\left(0\right) < {p_{{\rm{s,max}}}}$ and $p_\mathrm{s}^{\left( i \right)}\left(1\right) > {p_{{\rm{s,max}}}}$. Moreover, the intersection between $p_{\rm{s}}^{\left( 1 \right)}\left(\mathcal{C}_x\right)$ and $p_{\rm{s}}^{\left( 2 \right)}\left(\mathcal{C}_x\right)$ in the interested interval, if they are not identical, is computed from \eqref{intersect_r3}, which exists if $p_s^{\left( i \right)}\left( 0 \right) < p_s^{\left( j \right)}\left( 0 \right)$ and $p_s^{\left( i  \right)}\left( 1 \right) > p_s^{\left( j \right)}\left( 1 \right)$. 
\begin{figure*}
\begin{align}\label{intersect_r3}
{r^{\left( 3 \right)}} = \sqrt {1 - \frac{{4{\beta _1}{\beta _2}{{\cal I}_{{{\rm{s}}_1}}}{{\cal I}_{{{\rm{s}}_2}}}\left( {{\beta _1}{{\cal I}_{{{\rm{s}}_2}}}{\Psi _2}\left( {1,2} \right) - {\beta _2}{{\cal I}_{{{\rm{s}}_1}}}{\Psi _1}\left( {1,2} \right)} \right)\left( {{\beta _2}{{\cal I}_{{{\rm{s}}_1}}}{\Psi _2}\left( {1,2} \right){\Psi _1}\left( {2,2} \right) - {\beta _1}{{\cal I}_{{{\rm{s}}_2}}}{\Psi _1}\left( {1,2} \right){\Psi _2}\left( {2,2} \right)} \right)}}{{{{\left( {\beta _2^2{\cal I}_{{{\rm{s}}_1}}^2{\Psi _1}\left( {2,2} \right) - \beta _1^2{\cal I}_{{{\rm{s}}_2}}^2{\Psi _2}\left( {2,2} \right)} \right)}^2}}}}, 
\end{align}
\end{figure*}
After that, we define the interval boundaries points as ${\cal C}_ x^{(z)}$, where $z$ is an integer number in  $ [1,k+1]$, $k$ is the number of  distinct  intersection points, i.e. $k \in \{ 0,1,2,3 \}$, ${\cal C}_ x ^{(0)}=0$, ${\cal C}_x ^{(k+1)}=1$ and ${\cal C}_ x^{(1)}$, ${\cal C}_ x^{(2)}$ and ${\cal C}_ x^{(3)}$ are the ordered  distinct  intersection points (if exist). 

Hence, we can divide the optimization problem in \eqref{opt_prob_mod} into $(k+1)$ subproblems, where  each subproblem is defined in a specific range ${\cal C}_x ^{(z-1)} \le {{\cal C}_x} \le {\cal C}_x ^{(z )}$. Then, we compute the optimal solution that achieves the maximum SU rate. After obtaining the minimum of the three functions in \eqref{ps_min} in the interested interval, the $z^{\mathrm{th}}$ subproblem is formulated as
\begin{align}\label{{opt_subprob}}
& {\mathfrak{P}_z}:\mathop {\max }\limits_{{\cal C}{_x}} {R_{\rm{s}}}\left( {{\cal C}{_x}} \right)\nonumber \\ & \qquad \mathrm{s.\;t.} \hspace{0.25cm}{\cal C}_x^{\left( {z - 1} \right)} \le {{\cal C}_x} \le {\cal C}_x^{\left( z \right)}.
\end{align}
In order to solve ${\mathfrak{P}_z}$, there are two cases, either $p_{\rm{s}}=p_{\rm{s}}^{\left( i \right)}\left({{\cal C}_x}\right)$ or $p_{\rm{s}}={p_{{\rm{s,max}}}}$. If $p_{\rm{s}}= p_{\rm{s}}^{\left( i \right)}\left({{\cal C}_x}\right)$ in  \eqref{su_rate}, then ${R_{\rm{s}}}\left( {{\cal C}{_x}} \right)$ is  strictly increasing in $\mathcal{C}_x$ if and only if\footnote{See Appendix B for the proof.}
\begin{align}\label{improper_condition}
\frac{{{\beta _j}{\gamma _{\rm{s}}}{\Psi _i}\left( {1,2} \right)}}{{{{\cal I}_{{{\rm{s}}_j}}}\left( {\sum\nolimits_{i = 1}^2 {{p_i}{{\cal I}_{{{\rm{p}}_i}}}}  + 1} \right)}} >  - 1.
\end{align}
Hence, the optimal solution pair in this case is $ ( {p_{\rm{o}}^{(z)},{\cal C}_{\rm{o}}^{(z)}} )= (p_s^{\left( i \right)} ({\cal C}_ x ^{(z)} ), {\cal C}_x ^{(z)} )$.  Otherwise, it is a strictly decreasing function and hence,  $ ( {p_{\rm{o}}^{(z)},{\cal C}_{\rm{o}}^{(z)}} )= (p_s^{\left( i \right)} ({\cal C}_ x ^{(z-1)} ), {\cal C}_x ^{(z-1)} )$. Moreover, if $p_{\rm{s}}={p_{{\rm{s,max}}}}$ in \eqref{ps_min}, it is clear that ${R_{\rm{s}}}\left( {{\cal C}{_x}} \right)$ is strictly decreasing in $\mathcal{C}_x$,  hence, the optimal solution pair is $( {p_{\rm{o}}^{(z)},{\cal C}_{\rm{o}}^{(z)}} )= ({p_{{\rm{s,max}}}}, {\cal C}_ x^{(z-1)})$. At the end, we pick the optimal pair $( {p_{\rm{o}}^{(z)},{\cal C}_{\rm{o}}^{(z)}} )$ that has the maximum cost value. Based on the aforementioned analysis, we design Algorithm I to find the optimal solution pair $\left( {p_{\rm{s}}^*,{\cal C}_x^*} \right)$ that solves the IGS optimization problem.  

\floatname{algorithm}{}
\begin{algorithm} \label{alg1}
\renewcommand{\thealgorithm}{}
\newcommand{\tab}[1]{\hspace{.06\textwidth}\rlap{#1}}
\caption{\textbf{Algorithm I}}
\begin{algorithmic}[1]
\State \textbf{Input} $p_i$, ${ \gamma _{{{\mathrm{p}}_{_i}}}}$, ${{ {\cal I}}_{{{\rm{p}}_i}}}$, ${ \upsilon _{{{\mathrm{p}}_i}}}$, ${R_{0,{{\rm{p}}_i}}}$ ,  ${{ \mathcal I}_{{{\mathrm{s}}_i}}}$, $p_{\mathrm{s}}^{(0)} \hspace{-0.1cm}= {p_{{\rm{s,max}}}}$, ${\gamma _{\rm{s}}}$, ${\cal C}_ x^{(z)}$, and ${\cal C}_ x^{(z-1)}$
\State \textbf{Compute} ${{\cal I}_{{\rm{max,}}{{\rm{p}}_i}}}$ from \eqref{Imax}.
\If {$\left({\cal I}_{{\rm{max,}}{\rm{p}_i}} > {p_{_j}}\upsilon _{{\rm{p}}_j}\right)$}\vspace{0.1cm}
\For {$z=1:k+1$}
\State \textbf{Compute} $m = \mathop {\arg \min }\limits_{l \in \{0,1,2\}}   p_{\mathrm{s}}^{( l  )} \left(  \frac{{\cal C}_ x^{(z-1)} +{\cal C}_ x^{(z)}}{2} \right)$ 
\If{ $m =0$}\vspace{0.1cm}
\State $p_{\rm{o}}^{(z)}\leftarrow {p_{{\rm{s,max}}}}$,   $ \quad \mathcal{C}_\mathrm{o}^{(z)}\leftarrow {\cal C}_ x ^{(z-1)}$\vspace{0.1cm}
\ElsIf{ $\left(\frac{{{\beta _j}{\gamma _{\rm{s}}}{\Psi _m}\left( {1,2} \right)}}{{{{\cal I}_{{{\rm{s}}_j}}}\left( {\sum\nolimits_{i = 1}^2 {{p_i}{{\cal I}_{{{\rm{p}}_i}}}}  + 1} \right) }} >  - 1\right)$, $j\neq m$}\vspace{0.1cm}
\State $p_{\rm{o}}^{(z)}\leftarrow p_\mathrm{s}^{\left( m \right)}\left({\cal C}_x ^{(z)}\right)$, $\quad \mathcal{C}_{\rm{o}}^{(z)}\leftarrow {\cal C}_x ^{(z)}$
\Else {}
\State $p_{\rm{o}}^{(z)}\leftarrow p_{\mathrm{s}}^{\left( m \right)}\left({\cal C}_x ^{(z-1)}\right)$, $\quad\mathcal{C}_{\rm{o}}^{(z)}\leftarrow {\cal C}_x ^{(z-1)}$
\EndIf 
\EndFor
\State \textbf{Output} $\left( {p_{\rm{s}}^*,{\cal C}_x^*} \right) = \mathop {\arg \max }\limits_{p_{\rm{o}}^{(z)},\;{\cal C}_o^{(z)}} {R_{{\rm{s}}}}\left( {p_{\rm{o}}^{(z)},{\cal C}_{\rm{o}}^{(z)}} \right)$
\Else{}
\State \textbf{Output} $\left( {p_{\rm{s}}^*,{\cal C}_x^*} \right)=\left(0,0\right)$
\EndIf

\end{algorithmic}
\end{algorithm}

\section{Simulation Results}
In this section, we conduct some numerical examples which investigate the benefits of adopting IGS in improving the opportunities of the SU to share the spectrum with FD PU. The simulations are done over $10^5$ independent channel realizations except for $h_{ij}$ and $h_{ji}$ which are generated with a correlation coefficient of 0.95. Throughout the simulations, we use the following simulation parameters, unless otherwise specified. For the PU nodes, we assume ${R_{0,{{\rm{p}}_i}}}=1$ b/s/Hz with a maximum power budget $p_i=1 \; \mathrm{W}$. The communications channels are characterized as, ${\bar \gamma _{{{\mathrm{p}}_{_i}}}}={\bar \gamma _{{{\mathrm{p}}}}}=15$ dB, $\left(\bar {\cal I}_{{{\rm{p}}_1}},\bar {\cal I}_{{{\rm{p}}_2}}\right)=\left(5,8\right)$ dB, ${\bar \upsilon _{{{\mathrm{p}}_i}}}={\bar \upsilon _{{{\mathrm{p}}}}}=10$ dB. The SU is assumed to have a maximum power budget $p_{\rm{s,max}}=1 \; \mathrm{W}$. The SU channels parameters are  $\left(\bar {\cal I}_{{{\rm{s}}_1}},\bar {\cal I}_{{{\rm{s}}_2}}\right)=\left(20,10\right)$ dB and ${\bar \gamma _{{{\mathrm{s}}}}}=15$ dB. The PGS design is based on \eqref{proper_psi}. For the IGS design, we first get the distinct intersection points, if exist, using  \eqref{intersect_r1_r2} and \eqref{intersect_r3} and sort them in $\mathcal{C}_x^{z}$, then we apply Algorithm I to obtain the optimal pair $\left(p_{{\rm{s}}}^*,\mathcal{C}_x^*\right)$.
    
\begin{figure}[!t]
\centering
\includegraphics[width=9cm]{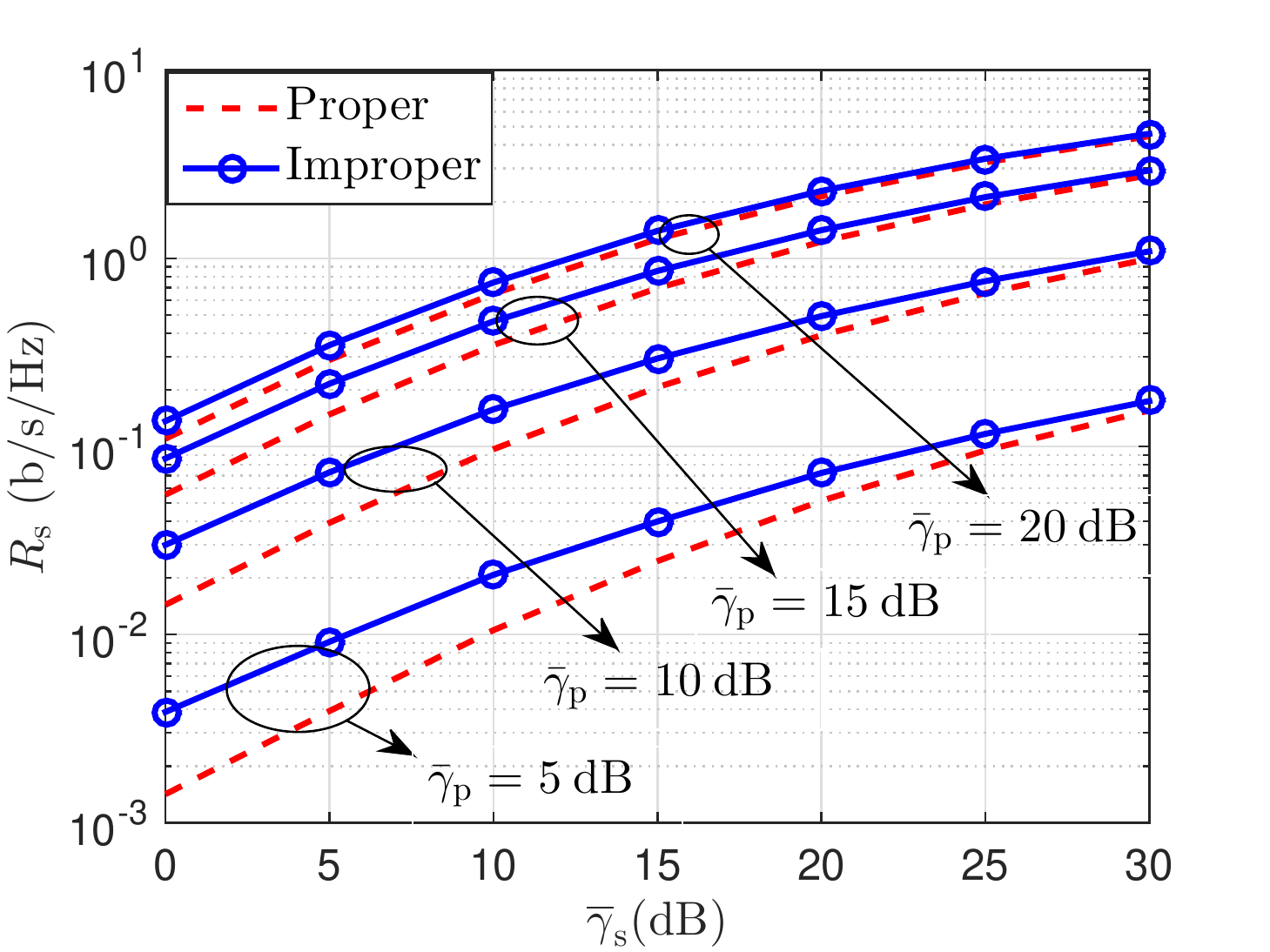}
\caption{SU achievable rate versus ${\bar \gamma _{{{\mathrm{s}}}}}$ for different ${\bar \gamma _{{{\mathrm{p}}}}}$ values.}
\label{SimEx1}
\end{figure}

\textit{\textbf{Example 1:}}
This example investigates the benefits of deploying IGS over the conventional PGS for spectrum sharing with FD PU and inspect the effects of the direct PU and SU channels on the performance of such systems. Fig. \ref{SimEx1} plots the achievable rate of the SU versus ${\bar \gamma _{{{\mathrm{s}}}}}$ for different values  of ${\bar \gamma _{{{\mathrm{p}}}}}$. Interestingly, the IGS scheme can boost the rate performance at small ${\bar \gamma _{{{\mathrm{s}}}}}$ achieving  a $1-4$ dB improvement over PGS.

At lower values of ${\bar \gamma _{{{\mathrm{p}}}}}$, the PU cannot afford more interference from the SU. Hence, PGS uses less power while IGS increases its transmitted power and  compensates for its interference impact on the PU by increasing the circularity coefficient $\mathcal{C}_x$. On the other hand, at higher values of ${\bar \gamma _{{{\mathrm{p}}}}}$, the PU has a bigger room for interference and hence, PGS can use the maximum budget and still satisfy the PU QoS. Thus IGS and PGS reduce to approximately the same solution. IGS can achieve better performance if the SU have more power budget as it is discussed in simulation example 3.   
         
\begin{figure}[!t]
\centering
\includegraphics[width=9cm]{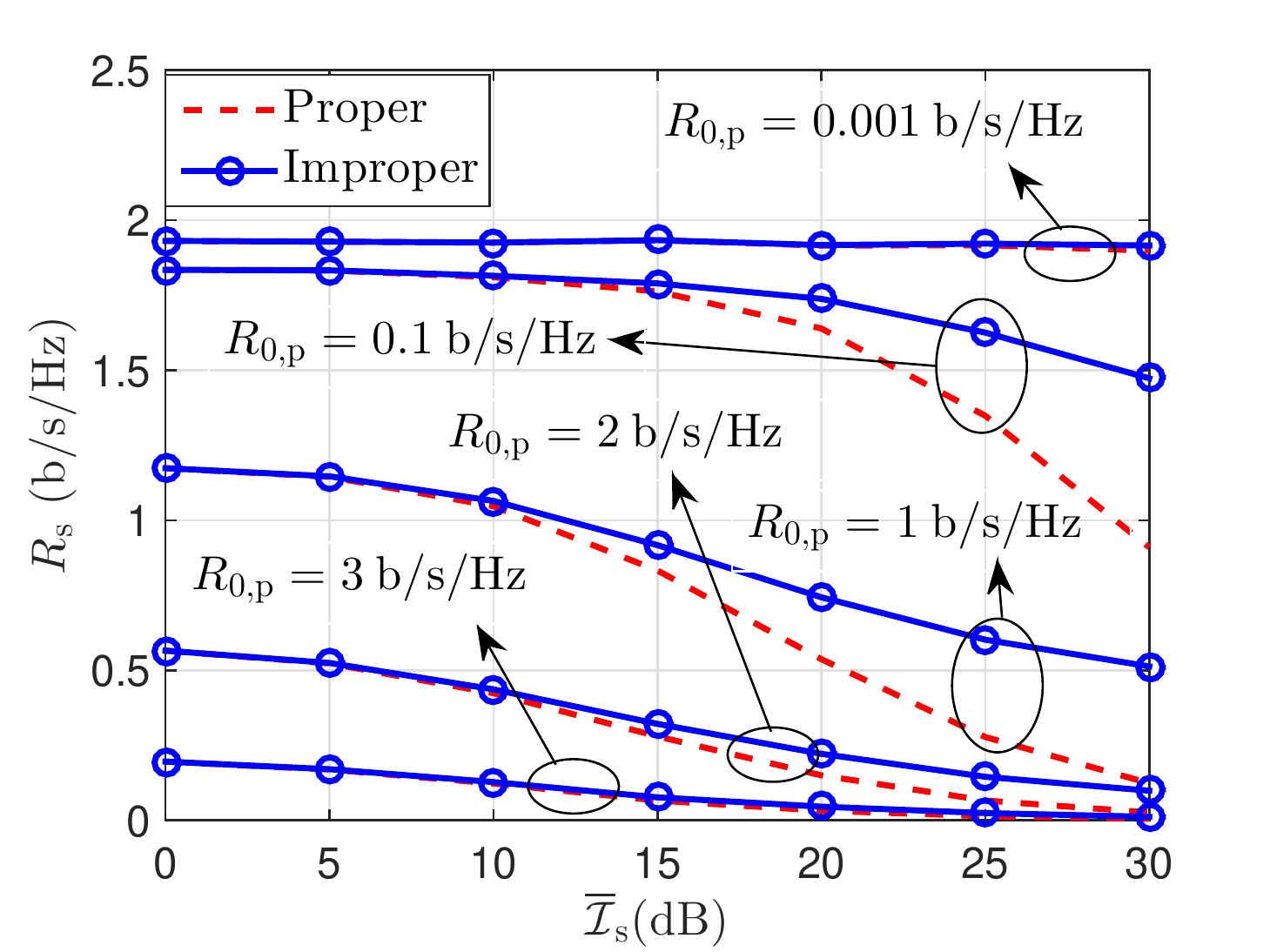}
\caption{SU achievable rate versus ${{\bar {\cal I}}_{{{\rm{s}}}}}$ for different PU rates ${R_{0,{{\rm{p}}}}}$.}
\label{SimEx2}
\end{figure}

\textit{\textbf{Example 2:}}
In this example, the effect of the SU interference channel to the PU and the minimum PU target rates are studied. For this purpose, we plot the SU achievable rate versus ${{\bar {\cal I}}_{{{\rm{s}}}}}$ for different PU rates ${R_{0,{{\rm{p}}}}}$ in Fig. \ref{SimEx2}, assuming that ${{\bar {\cal I}}_{{{\rm{s}}_i}}}={{\bar {\cal I}}_{{{\rm{s}}}}}$ and ${R_{0,{{\rm{p}}_i}}}={R_{0,{{\rm{p}}}}}$. At lower values of ${{\bar {\cal I}}_{{{\rm{s}}}}}$, there are no significant benefits of employing the IGS scheme when compared with the PGS scheme. This is due to the weak interference channel, which permits increasing the SU power, thus, the IGS solution reduces to the PGS one. On the other hand, when $\bar {{\cal I}}_{{{\rm{s}}}}$ become stronger, PGS tends to use less power to meet the PU QoS while IGS uses more power to improve its performance and relieve its interference impact on the PU through increasing the circularity coefficient.

At very low PU rates ${R_{0,{{\rm{p}}}}}$, both IGS and PGS use approximately the maximum power budget as the PU QoS requirements are relaxed. As ${R_{0,{{\rm{p}}}}}$ increases, the allowable interference room in the PU network decreases, which forces the PGS to reduce the transmitted power. On the other hand, IGS has the ability to use more power to increase the SU rate performance. At high PU rates, the PU nodes cannot afford more interference from the SU and thus both IGS and PGS use less power to meet the PU QoS requirements. Therefore, the performance gain of deploying the IGS  scheme decreases till a certain value for ${R_{0,{{\rm{p}}}}}$ at which both PGS and IGS schemes converges almost to the same solution. 

\begin{figure}[!t]
\centering
\includegraphics[width=9cm]{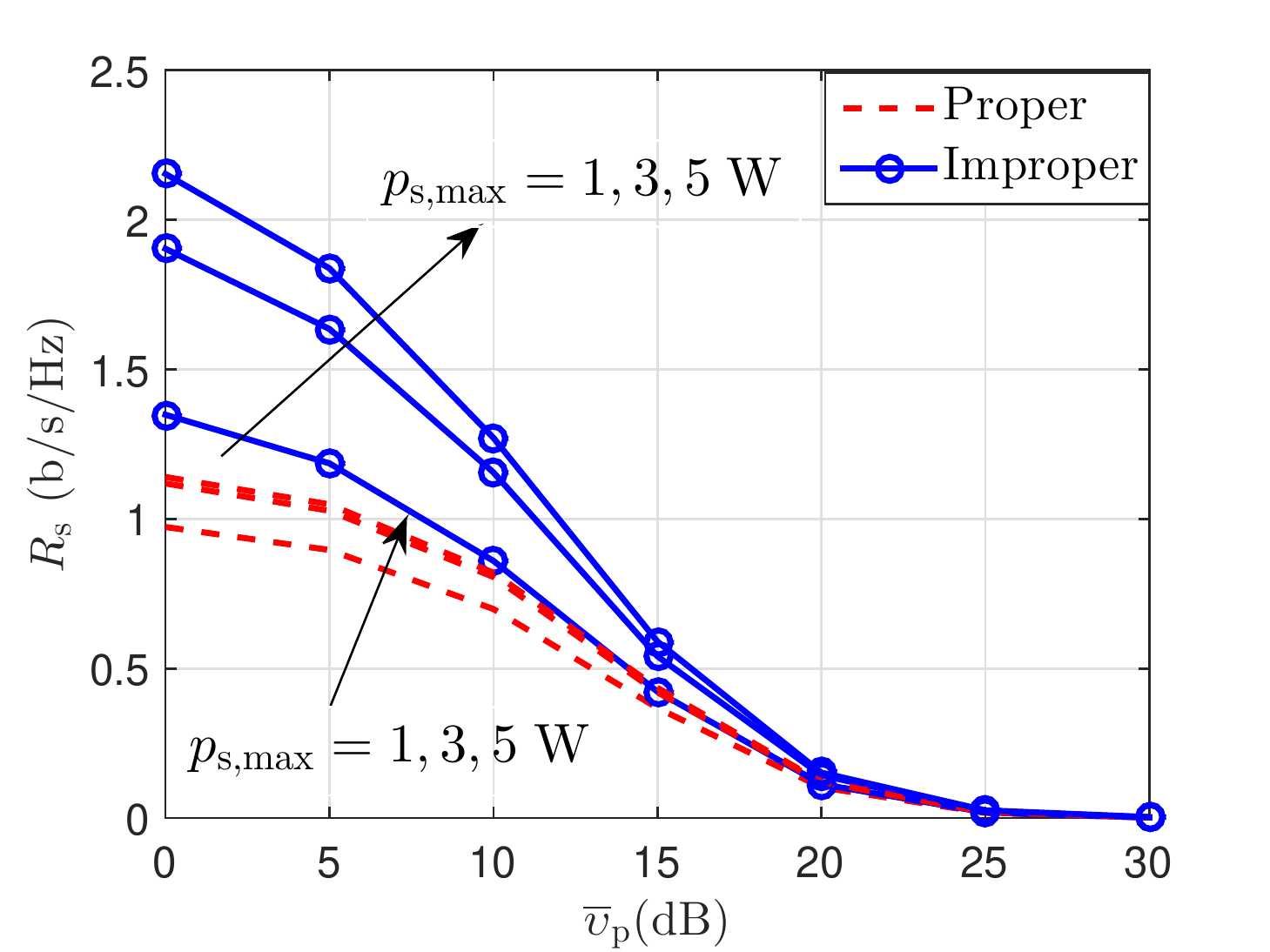}
\caption{SU achievable rate versus ${\bar \upsilon _{{{\mathrm{p}}}}}$ for different maximum power budget ${p_{{\rm{s,max}}}}$.}
\label{SimEx3}
\end{figure}

\textit{\textbf{Example 3:}}
This example investigates the impact of RSI in limiting the underlay CR operation and compares between its effect on both IGS and PGS based systems. To this end, we plot $R_{\mathrm{s}}$ versus ${\bar \upsilon _{{{\mathrm{p}}}}}$ for different  ${p_{{\rm{s,max}}}}$ in Fig. \ref{SimEx3}. We notice that IGS achieves better performance  than PGS until a certain value of the RSI CNR at which the PU is overwhelmed by the RSI and both signaling schemes fail to operate properly. 

Furthermore, PGS system cannot get more benefits from increasing the SU power budget while the IGS tends to use the total budget efficiently and relieve the interference effect on PU by increasing $\mathcal{C}_x$, which reduces the interference impact as can be seen from \eqref{pu_rate_not_simplified}.

\section{Conclusion}
In this paper, we consider the challenging spectrum sharing scenario when the PU uses in-band FD paradigm. We employ IGS to enhance the opportunities of sharing the resources of the proposed system. The SU signal parameters is optimized to maximize the SU achievable rate subject to a minimum PU target rate constraint and a maximum SU power budget. We develop a simple algorithm that optimally computes the SU power and circularity coefficient. The numerical results show that the SU can achieve a significant performance gain through adopting IGS. The main advantages of the proposed scheme are for week PU direct channels and/or for strong SU interference channels to PU. PGS tends to use less transmitted power, while IGS uses more power and compensate for its interference signature through increasing the circularity coefficient. Thus, IGS gets benefits from increasing the SU power budget and uses it efficiently unlike PGS.

\section*{Appendix A}
In this appendix, we prove that $p_\mathrm{s}^{\left( i \right)}\left( {{\cal C}{_x}} \right)$ is strictly increasing in ${\cal C}_x$ over the interval $0 < {{\cal C}_x} < 1$ by taking the first derivative as follows
\begin{align}\label{ps_inc_proof}
\frac{{dp_{\rm{s}}^{\left( i \right)}\left( {{\cal C}{_x}} \right)}}{{d{{\cal C}_x}}} =& \frac{{{{\cal C}_x}{\beta _j}\left|{\Psi _i}\left( {1,2} \right)\right|}}{{{{\bar {\cal I}}_{{{\rm{s}}_j}}}{{\left( {1 - {\cal C}_x^2} \right)}^2}\sqrt {1 + {\Omega _i}} }}\Bigg[ \left( {{\Omega _i} + 2} \right) + \nonumber \\
&{\mathop{\rm sgn}} \left\{ {{\Psi _i}\left( {1,2} \right)} \right\}\sqrt {{{\left( {{\Omega _i} + 2} \right)}^2} - \Omega _i^2}  \Bigg],
\end{align}
where ${\Omega _i} = \frac{{\left( {1 - {\cal C}_x^2} \right){\Psi _i}\left( {2,2} \right)}}{{\Psi _i^2\left( {1,2} \right)}}>0$ and the sign function, ${\mathop{\rm sgn}} \left\{ x \right\}$, is defined as
\begin{align}
{\mathop{\rm sgn}} \left\{ x \right\} = \left\{ {\begin{array}{*{20}{l}}
{- 1}&{x < 0},\\
0&{x = 0},\\
1&{x > 0}.
\end{array}} \right.
\end{align}
Hence, \eqref{ps_inc_proof} is always positive and this concludes the proof.
\section*{Appendix B}
In this appendix, we derive the conditions in \eqref{improper_condition} over the interested interval $0 < {{\cal C}_x} < 1$. From \eqref{ps_improper_quadraric}, we obtain
\begin{align}\label{ps_squared}
{\left( {p_{\rm{s}}^{\left( i \right)}\left( {{{\cal C}_x}} \right)} \right)^2} \le \frac{{\left( {\beta _j^2{\Psi _i}\left( {2,2} \right) + 2p_{\rm{s}}^{\left( i \right)}\left( {{{\cal C}_x}} \right){\beta _j}{{\cal I}_{{{\rm{s}}_j}}}{\Psi _i}\left( {1,2} \right)} \right)}}{{{\cal I}_{{{\rm{s}}_j}}^2\left( {1 - {\cal C}_x^2} \right)}}.
\end{align}
By substituting \eqref{ps_squared} in \eqref{su_rate}, we get
\begin{align}\label{su_rate_proof}
{R_{\rm{s}}}\left( {\cal C}{_x} \right) = \frac{1}{2}{\log _2}\Bigg(& \frac{{2{\gamma _{\rm{s}}}{p_{\rm{s}}^{\left( i \right)}\left( {{{\cal C}_x}} \right)}}}{\Delta }\left( {1 + \frac{{{\beta _j}{\gamma _{\rm{s}}}{\Psi _i}\left( {1,2} \right)}}{{{{\cal I}_{{{\rm{s}}_j}}}\Delta }}} \right) + \nonumber \\
&\frac{{\beta _j^2{\Psi _i}\left( {2,2} \right)\gamma _{\rm{s}}^2}}{{{\cal I}_{{{\rm{s}}_j}}^2{\Delta ^2}}} + 1 \Bigg),
\end{align} 
where $\Delta  = \sum\nolimits_{i = 1}^2 {{p_i}{{\cal I}_{{{\rm{p}}_i}}}}  + 1>0$. We show in Appendix A that $p_\mathrm{s}^{\left( i \right)}\left( {{\cal C}{_x}} \right)$, is a strictly increasing function in ${\cal C}_x$ over the interval $0 < {{\cal C}_x} < 1$. Moreover, $p_\mathrm{s}^{\left( i \right)}\left( {{\cal C}{_x}} \right)$ represents the only dependency on ${\cal C}_x$ in \eqref{su_rate_proof} and since the logarithmic function ${\log _2}\left( x \right)$ is strictly increasing for $x>0$, ${R_{\rm{s}}}\left( {\cal C}{_x} \right)$ is strictly increasing in ${{\cal C}_x}$ when the slope of the linear term inside the logarithm with respect to  $p_\mathrm{s}^{\left( i \right)}\left( {{\cal C}{_x}} \right)$ is positive, which results in the following condition   
\begin{align}
{1 + \frac{{{\beta _j}{\gamma _{\rm{s}}}{\Psi _i}\left( {1,2} \right)}}{{{{\cal I}_{{{\rm{s}}_j}}}\Delta }}}>0.
\end{align}
This gives the conditions in \eqref{improper_condition} and concludes the proof.

\bibliographystyle{IEEEtran}

\bibliography{IEEEabrv,mgaafar_April_2015}

\vfill

\end{document}